\documentclass{article}
\usepackage[T1]{fontenc}

\makeatletter

\newcommand{\LyX}{L\kern-.1667em\lower.25em\hbox{Y}\kern-.125emX\@}

\newcommand{\lyxaddress}[1]{
  \par {\raggedright #1 
  \vspace{1.4em}
  \noindent\par}
}

\makeatother

\begin{document}

\title{Biological Networks}

\author{Indrani Bose}

\maketitle

\lyxaddress{Department of Physics, Bose Institute, 93/1, A.P.C. Road, Calcutta-700009,
India}

\begin{abstract}
In this review, we give an introduction to the structural and functional properties
of the biological networks. We focus on three major themes: topology of complex
biological networks like the metabolic and protein-protein interaction networks,
nonlinear dynamics in gene regulatory networks and in particular the design
of synthetic genetic networks using the concepts and techniques of nonlinear
physics and lastly the effect of stochasticity on the dynamics. The examples
chosen illustrate the usefulness of interdisciplinary approaches in the study
of biological networks.
\end{abstract}

\section{Introduction}

Networks are widely prevalent in all spheres of life \cite{1,2,3}. A network
of acquaintances is the simplest example one can think of. Social, economic
and political networks of various kinds are part of human society. The internet,
a network of information resources, plays a vital role in the gathering, sharing
and transmission of information. A network consists of nodes connected by links.
Figure 1 shows the example of a network in which the solid circles denote the
nodes and the solid lines the links. Some examples of real life networks are
as follows: in a network describing an electrical power grid, the generators,
transformers and substations are the nodes and the high-voltage transmission
lines connecting them the links. In the World Wide Web (WWW), the documents/pages
constitute the nodes. These are connected to other documents/pages through links.
In a collaboration graph of movie actors, the nodes represent the actors. Two
actors are connected by a link if they appear in the same movie. In a citation
network, the nodes are the papers published in refereed journals. A paper is
linked to all the other papers it cites. Cellular processes are controlled by
various types of biochemical networks. A metabolic network \cite{4} controls
the processes which generate mass and energy from nutritive matter. The nodes
in such a network are the substrates such as ATP, ADP and \( H_{2}O \). Two
substrates are connected by a link if both of them participate in the same biochemical
reaction. Traditional cell biology assigns specific functional roles to individual
proteins, such as catalysts, signalling molecules and constituents of cellular
matter. In the post-genomic era, there is an increasing emphasis on understanding
the functions of proteins as parts of an interacting network and also on the
collective, emergent properties of the network. In a protein-protein interaction
network \cite{5}, the nodes represent the proteins. A link exists between two
nodes if the corresponding proteins have a direct physical interaction.

The networks discussed above have complex topology. Spectacular advances in
computerisation of data acquisition (the Human Genome Project is a prime example)
have made it possible to construct large databases which contain information
on the topology of real life networks. The advent of powerful computers has
given rise to extensive investigations of networks containing millions of nodes.
The interesting fact emerging out of these studies is that biological networks
share common topological features with non-biological networks. There appears
to be a general blueprint for the large scale organisation of several of these
networks. In Section 2 of this review, we discuss two types of biological networks,
namely, the metabolic networks of several organisms and the protein-protein
interaction networks associated with the yeast S. cerevisiae \cite{6} and the
human gastric pathogen H. Pylori \cite{7}. The major topological features of
these networks are described and the similarity in the design principles of
large-scale biological and non-biological networks pointed out.

Gene regulatory networks are the most significant examples of biological networks.
Gene expression and regulation are the central activities of a living cell \cite{8}.
Genes are fragments of \( DNA \) molecules and determine the structure of functional
molecules like \( RNAs \) and proteins. In each cell, at any instant of time,
only a subset of genes present is active in directing \( RNA \) / protein synthesis.
The gene expression is ``on'' in such a case. The information present in the
gene is expressed through the processes of transcription and translation. During
transcription, the sequence along one of the strands of the \( DNA \) molecule
is copied onto a \( RNA \) molecule (\( mRNA \) ). The sequence of the \( mRNA \)
molecule is then translated into the sequence of amino acids, which determines
the functional nature of the protein molecule produced. In a gene regulatory
network, the protein encoded by one gene can regulate the expression of other
genes. These genes in turn produce new regulatory proteins which control still
other genes. A protein may also regulate its own level of production through
an autoregulatory feedback process. The occurrence of cell differentiation,
when an organism grows from its embryonic stage, depends upon the selective
switching on of gene expression in individual cells. All these cells have identical
sets of genes but follow different developmental pathways depending upon the
patterns of gene expression in the cells. Thus distinct types of cells such
as hair and skin cells are obtained. Gene expression is also regulated in metabolism
and progression through the cell cycle as well as in responses to external signals.
Infected cells can multiply because the expression of certain genes is ``on''
in these cells whereas in normal cells the expression of the same genes is ``off''.

Despite a vast amount of experimental data, the complex dynamical processes
involved in gene regulation are not fully understood as yet. A large number
of theoretical studies has been undertaken \cite{9} but only a few of these
make quantitative predictions in agreement with experimental results. Two key
concepts which emerge out of the theoretical studies are: nonlinearity of the
network dynamics and the role of stochasticity in gene expression and regulation
\cite{10}. The variables of interest in the network dynamics are the concentrations
of the \( mRNAs \), proteins and other biomolecules within the cell. The rate
of change in the concentration of a biomolecular species is a nonlinear function
of the other variables. The dynamics is governed by a set of coupled non-linear
differential equations which in most cases are solved numerically. Let \( U \)
be the concentration of, say, a particular type of protein in the cell. The
rate of change of \( U \) is given by

\( \frac{dU}{dt}= \) (Production - Loss/Decay ) of \( U \) per unit time.\\
The production term is a nonlinear function of the other concentration variables
and the loss term is usually proportional to \( U \). In Section 3 of this
review, we briefly describe the major features of nonlinearity in the dynamics
of gene networks. There is currently a significant emerging trend to utilise
the concepts and techniques of nonlinear physics in the actual construction
of synthetic gene regulatory networks with a variety of applications. In Section
3, a specific example of this, namely, the genetic toggle switch \cite{11}
will be given. 

The biochemical rate equations which govern the dynamics of gene regulatory
networks are deterministic in nature. Many molecules associated with the networks
have low intracellular concentrations and consequently fluctuations in reaction
rates are considerably large. Gene expression involves a series of biochemical
reactions and due to stochastic fluctuations in the reaction rates, proteins
are produced in short bursts at random time intervals. In the last few years,
there is an increasing realization that stochasticity plays a significant role
in biological processes \cite{10,12}. To give an example, consider the situation
in which two independently produced regulatory proteins A and B are in competition
to control a developmental switch that selects between two pathways depending
on which protein wins. The protein concentrations have to reach effective levels
in order to activate the switch. Due to stochastic fluctuations, the amounts
of proteins A and B produced as a function of time can vary widely from cell
to cell. In some cells, protein A reaches the effective level first and activates
the developmental switch along one specific pathway. In the other cells, protein
B takes control and the other pathway is activated. Thus even a clonal cell
population exhibits phenotypic variations as the cells follow different developmental
pathways. Environmental signals can bias the probabilities of path choice in
a regulatory circuit. Organisms make use of this mechanism to increase the probability
of survival in a hostile environment. Cells often utilise fluctuations (noise)
to randomize choices of developmental pathways when such randomization is desirable
for the survival and growth of the organism. A well-known example is that of
the phage \( \lambda  \) lysis-lysogeny network \cite{13}. The bacterial E.coli
cells, when infected by the virus phage \( \lambda  \) , can follow two developmental
pathways: lysis and lysogeny. In the lysogenic state, the infection is dormant.
Phage \( \lambda  \) is inert and integrated into the host cell's chromosome.
It replicates along with the bacterial DNA and each new cell contains the dormant
phage. In the lytic state, the infection proliferates. The viral DNA replicates
using the host cell machinery giving rise to a large number of progeny phage.
These in turn lyse or burst the host bacterium cell and the infection spreads
to more cells. Again, due to stochastic fluctuations, the cell population divides
into two subpopulations: lysogenic and lytic. The selection of a developmental
pathway after the host cell is infected is not deterministic but probabilistic.
In Section 4 of this review, the effect of stochasticity on the dynamics of
the \( \lambda  \)-phage network will be briefly discussed. The network illustrates
the competitive control of a developmental switch by two regulatory proteins.

As already mentioned before, gene expression/regulation involves several biochemical
reactions with appreciable stochastic fluctuations. Gillespie \cite{14,15}
has proposed a Monte Carlo simulation algorithm to describe the kinetics of
coupled stochastic reactions. This method is physically more rigorous than the
conventional differential equation approach. The inherent assumption in the
latter method is that the temporal changes in the concentrations of reacting
molecules are both continuous and deterministic. The assumption is not true
if the concentrations are small and the reaction rates slow or if the system
undergoes large, rapid and discrete transitions. In the Gillespie algorithm,
changes in the numbers of the reacting molecules occur in integral numbers brought
about by random, distinct reaction events. The Gillespie algorithm is described
in detail in Section 4 and some illustrative examples are given. Recent experiments
at the level of a single cell have shown that gene expression occurs in abrupt
stochastic bursts \cite{16,17,18,19}. Further, in an ensemble of cells, the
levels of proteins produced have a bimodal distribution. In a large fraction
of cells, the gene expression is either off or has a high value. We have proposed
a model of gene expression the essential features of which are stochasticity
and cooperative binding of RNA polymerase, the molecule responsible for transcription
\cite{20}. The model can reproduce the bimodality observed in experiments.
We include a description of the model in Section 4 to give an additional example
of the effect of stochasticity on gene regulation. Section 5 of the review contains
concluding remarks. 

The emphasis in this review is on recent studies of biological networks. There
are a number of exhaustive reviews and books on earlier work. The three major
themes that several recent studies focus on are: topology of networks, nonlinear
dynamics and its consequences and the role of stochasticity in biological processes.
The present review is meant to be an introduction to these themes and to highlight
the fact that interdisciplinary approaches are essential to develop an integrated
understanding of biological networks.

\section{Topology of complex networks}

Many real life networks have a complex structure. The mathematicians Erd\"{o}s
and R\'{e}nyi \cite{21} were the first to propose a model of a complex network
known as a random graph. One starts with \( N \) nodes and connects every pair
of nodes with probability \( p \). The graph thus has approximately \( \frac{pN(N-1)}{2} \)
links distributed in a random manner. Studies of real life networks, however,
reveal that these cannot be described as random graphs. This distinction is
possible on the basis of quantitative measurements of certain topological features
which we define below. Several complex networks including the random graph are
described as small world networks \cite{1,2,3,22}. The small world idea implies
that though the networks are large in size (the number of nodes in a network
is a measure of its size), any pair of nodes can be connected by a short path.
The distance between two nodes is given by the number of links along the shortest
path connecting the nodes. In Figure 1, the distance between the nodes A and
B is three. The diameter of the network, also known as the average path length
l, is the average of the distances between all pairs of nodes. The global population
is huge but still we live in a small world as any random pair of individuals
are connected to each other through a short path of intermediate acquaintances.
This was first established by Stanley Milgram \cite{23} who found out that
the average path length of intermediate acquaintances is six. In a small world
network, the diameter scales as the logarithm of the number of nodes.

The second measurable topological feature of a complex network is its degree
distribution \cite{1,2,3}. The number of links by which a node is connected
to the other nodes varies from node to node. Let \( P(k) \) be the probability
that a randomly selected node has exactly \( k \) links. Equivalently, \( P(k) \)
is the fraction of nodes, on an average, which has exactly \( k \) links. One
can define an average degree \( \left\langle k\right\rangle  \) of the network,
the degree of a node being the number of links attached to the node. In a random
graph, the links are established randomly and most of the nodes have degrees
close to \( \left\langle k\right\rangle  \) . The degree distribution \( P(k) \)
vs. \( k \) is Poissonian. It is strongly peaked at \( k=\left\langle k\right\rangle  \)
and has an exponential decay for large k, i.e., \( P(k)\sim e^{-k} \) for \( k\gg \left\langle k\right\rangle  \)
and \( k\ll \left\langle k\right\rangle  \) . In many real life networks, the
degree distribution \( P(k) \) has no well-defined peak but has a power-law
distribution

\begin{equation}
\label{1}
P(k)\sim k^{-\gamma }
\end{equation}
 where the exponent \( \gamma  \) is a numerical constant. Such networks are
known as scale-free networks because they are not tied to a specific scale.
\( P(k) \) has a finite value over a wide range of \( k \) values. The power-law
form of the degree distribution implies that the networks are extremely inhomogeneous
unlike in the case of a random graph. In a scale- free network, there are many
nodes with few links and a few nodes with many links. The highly connected nodes
play a key role in the functionality of the network. Both the random graph and
the scale-free networks are small world networks. 

The third topological quantity which is measurable is known as the clustering
coefficient \cite{2,22}. The coefficient is a measure of the tendency of the
nodes of the network towards clustering. In a social network, the individuals
are the nodes and two nodes are connected by a link if the individuals are acquainted
with each other. In such a network, one's friend's friends are also likely to
be one's friends giving rise to a clustering of acquaintances. The clustering
coefficient is defined in the following manner. Let us select a specific node
\( i \) in the network which is connected by \( k_{i} \) links to \( k_{i} \)
other nodes. If these first neighbours are all part of a cluster, there would
be \( \frac{k_{i}(k_{i}-1)}{2} \) links between them. The clustering coefficient
\( C_{i} \) of node i is given by

\begin{equation}
\label{2}
C_{i}=\frac{2E_{i}}{k_{i}(k_{i}-1)}
\end{equation}
 where \( E_{i} \) is the number of actual links which exist between the \( k_{i} \)
nodes. The clustering coefficient C of the whole network is obtained by taking
an average over all the \( C_{i} \) values. The utility of the clustering coefficient
is demonstrated in the following example. The neural network of the nematode
worm C. Elegans is small in size \cite{2,24}. The number of neurons which constitute
the nodes of the network is 282. A link exists between two nodes if the neurons
are connected by either a synapse or a gap junction. The average degree of the
network is \( \left\langle k\right\rangle = \)14. Now consider a random graph
of the same size and average degree. The average path lengths for the neural
network and the random graph are similar, 2.65 and 2.25 respectively. Is the
neural network then a random graph? The answer is no as the clustering coefficient
of the former has the value 0.28 which is much larger than the value 0.05 in
the case of the latter network. Examples of real life networks which are scale-free
are \cite{2,25}: the collaboration graph of movie actors (size \( N \) of
the network = 212 250 nodes, average degree \( \left\langle k\right\rangle = \)
28.78, the exponent \( \gamma  \) in Eq.(1) is \( \gamma  \) = 2.3 ), the
WWW (\( N= \) 325 729, \( \left\langle k\right\rangle = \) 5.46, \( \gamma = \)
2.1) and the network of citations (\( N= \) 783 339 papers, \( \left\langle k\right\rangle = \)
8.57, \( \gamma = \) 3). The results are obtained from available databases.
A more comprehensive and up to date list of networks is given in Ref. \cite{2}.

We now discuss some complex, biological networks. Recently, Jeong et al \cite{4}
have systematically investigated the topological properties of the core metabolic
networks of 43 different organisms representing all the three domains of life.
The data on these organisms are available in the WIT (What Is There) database.
As already mentioned in the Introduction, the nodes of the metabolic network
are the different substrates. Two substrates are connected by a link if they
participate in the same biochemical reaction. The metabolic networks have different
sizes, the less complex organisms having smaller sizes. There is considerable
variation in the individual constituents and the pathways of the networks. Yet
they display identical topological scaling properties which resemble those of
complex non-biological networks. The metabolic networks have been found to belong
to the class of scale-free networks. The probability that a substrate participates
in k reactions has a power-law distribution. The links in a metabolic network
are directed as many biochemical reactions are preferentially catalysed in one
direction. For each node, one has to distinguish between incoming and outgoing
links. Correspondingly, there are two exponents \( \gamma _{in} \) and \( \gamma _{out} \)
. The exponents turn out to have the same value of 2.2. 

In the metabolic network, the distance between two substrates is given by the
number of links (reactions) in the shortest biochemical pathway connecting the
two substrates. A surprising result obtained by Jeong et al is that the diameter
of the metabolic network is the same for all the 43 organisms, i.e., it does
not depend upon the number of substrates (nodes) belonging to the network. This
is counterintuitive and only possible if with increasing organism complexity
pre-existing individual substrates are increasingly connected in order to maintain
a more or less constant network diameter. In support of this conjecture, Jeong
et al found that the average number of reactions in which a certain substrate
participates increases as the number of substrates in the organism increases.
Conservation of the network diameter may be favourable for the survival and
growth of an organism. A larger diameter would possibly diminish the organism's
ability to respond to changes in an efficient manner. 

The scale-free character of the metabolic network implies that a few hubs which
are highly connected play a dominant role in the functioning of the network.
On sequential removal of these highly-connected nodes, the network diameter
rises sharply and ultimately the network disintegrates into isolated fragments.
On the other hand, the network diameter does not change appreciably when the
nodes with a few links are removed from the network. Scale-free networks, in
general, are robust against random mutations/errors but vulnerable to attacks
targeted at highly connected nodes. Complex communication networks are surprisingly
robust, local failures rarely hamper the global transmission of information.
Organisms can grow and survive in hostile environments due to the error tolerance
of the underlying metabolic network. A random graph is not as robust against
random mutations/errors. Mutagenesis studies in-silico and in-vivo \cite{26}
have established the remarkable error tolerance of the metabolic network of
E.coli on removing a large number of metabolic enzymes. Jeong et al, in their
study of the metabolic networks of organisms found that only \( \sim 4\% \)
of all the substrates present in all the 43 organisms are present in all the
species. The striking fact is that the small number of substrates, common to
all species, turn out to be the most highly connected ones. On the other hand,
there are species-specific differences in the case of less-connected substrates. 

Jeong et al in a separate study \cite{5} have investigated the protein-protein
interaction network of the yeast S.cerevisiae. The network has 1870 proteins
as nodes which are linked by 2240 direct physical interactions identified mostly
by systematic two-hybrid experiments. Actual measurements show that the probability
\( P(k) \) that a given yeast protein interacts with \( k \) other yeast proteins
has a power-law distribution with an exponential cutoff at \( k_{c}= \) 20.

\begin{equation}
\label{3}
P(k)\sim (k+k_{0})^{-\gamma }e^{-\frac{(k+k_{0})}{k_{c}}}
\end{equation}
 with \( k_{0}= \) 1 and \( \gamma = \) 2.4. The protein-protein interaction
network of the bacterium H. Pylori \cite{7} displays similar topology. For
the metabolic networks, the exponent \( \gamma  \) has the value 2.2. The value
of \( \gamma  \) falls in the range 2.0-2.5 for many scale-free networks. Like
the metabolic network and other scale-free networks, the protein-protein interaction
network is found to be immune to random mutations. The removal of highly connected
nodes may, however, disrupt the network function. The protein product of the
p53 tumor-suppressor gene is one of the most highly connected proteins found
in human cells. Mutations of p53 gene affect cellular functions severely from
a biomedical point of view.

In fact, Jeong et al's study on the protein-protein interaction network in yeast
shows that proteins with five or lesser number of links constitute \( \sim  \)
93\% of the total number of proteins but only \( \sim  \) 21\% of them are
essential so that the removal of such proteins proves to be lethal. In contrast,
only \( \sim  \) 0.7\% of the total number of proteins have more than 15 links
but single deletion of \( \sim  \) 62\% of these severely affects the functioning
of the network. It is possible that the proteins which constitute the highly
connected nodes in a network share common structural features. These features
favour the binding of many different types of proteins to the proteins in question.
The scale-free character of both the metabolic and protein-protein interaction
networks suggests the evolutionary selection of a common large scale structure
of biological networks. Studies of other biological networks are expected to
provide further evidence for this idea.

\section{Nonlinear dynamics}

The dynamics of gene regulatory networks are described by coupled nonlinear
p.d.e.'s which can be collectively represented as

\begin{equation}
\label{4}
\frac{dX(t)}{dt}=f(X,R)
\end{equation}
 where \( X(t) \) is the \( N \) -component state vector \( (X_{1}(t),...,X_{N}(t)) \)
and \( f \) is a set of nonlinear functions \( f_{1}(X,R),....,f_{N}(X,R) \)
. There are thus N coupled p.d.e.'s and an individual p.d.e. is of the form

\begin{equation}
\label{5}
\frac{dX_{i}(t)}{dt}=f_{i}(X_{1},....X_{N},R)
\end{equation}
There are in total N species of biochemical molecules participating in M reactions.
\( X_{i}(t) \) \( (i=1,...,N) \) represents the concentration of the ith molecular
species at time \( t \) . R represents a set of control parameters. The functions
\( f_{i}'s \) are nonlinear functions of the \( X_{i}'s \) and the specific
forms of the functions are determined by the structures and rate constants of
the M chemical reactions. As an example consider the set of reactions

\begin{equation}
\label{6}
P\rightarrow A
\end{equation}

\begin{equation}
\label{7}
A\rightarrow B
\end{equation}
 
\begin{equation}
\label{8}
A+2B\rightarrow 3B
\end{equation}
 
\begin{equation}
\label{9}
B\rightarrow C
\end{equation}
 The reactions represent the conversion of the precursor species \( P \) into
a final product \( C \) via a sequence of four reactions involving two intermediates
\( A \) and \( B \) . The third reaction is autocatalytic as \( B \) catalyses
its own production. The second reaction represents the uncatalysed conversion
of \( A \) to \( B \) and the last reaction shows that the catalyst \( B \)
decays into the product \( C \) . The equations are assumed to be irreversible.
Also, the concentration of the reactant \( P \) is assumed to be constant over
a reasonable period of time. This is possible if the initial concentration of
\( P \) is large. Let \( p_{0} \) (constant), \( a \) and \( b \) denote
the concentrations of the molecular species \( P \) , \( A \) and \( B \)
. The decay product \( C \) does not participate in any further reaction and
so does not influence the chemical kinetics. The rates of the four successive
chemical reactions are  \( k_{0}p_{0} \) , \( k_{u}a \) , \( k_{c}ab^{2} \)
and \( k_{d}b \) respectively where \( k_{0} \) , \( k_{u} \) , \( k_{c} \)
, \( k_{d} \) are the rate constants. The equations governing the chemical
kinetics are

\begin{equation}
\label{10}
\frac{da}{dt}=k_{0}p_{0}-k_{c}ab^{2}-k_{u}a
\end{equation}
 
\begin{equation}
\label{11}
\frac{db}{dt}=k_{u}a+k_{c}ab^{2}-k_{d}b
\end{equation}
 In the general scheme of p.d.e.'s shown in Eq. (5), \( N \) = 2, i.e., there
are two molecular species \( A \) and \( B \) participating in \( M \) =
4 chemical reactions. \( X_{1}=a \) and \( X_{2}=b \) are the concentrations
of the moleculeas \( A \) and \( B \). The r.h.s.'s of Eqs. (10) and (11)
are the nonlinear functions \( f_{1}(X_{1},X_{2}) \) and \( f_{2}(X_{1},X_{2}), \)
the nonlinearity arising from the autocatalytic term \( k_{c}ab^{2} \) . The
rate constants together with \( p_{0} \) constitute the control parameters
\( R \). 

In the general case, imagine an abstract \( N \) - dimensional state space
with axes \( X_{1},....,X_{N} \) . The state of the system at any instant of
time, say \( t_{0} \), is given by the \( N \) -component state vector \( X(t_{0}) \).
In the state space, this state is represented by a single point. The time evolution
of the system gives rise to a trajectory in the state space. The trajectory
may end up at a fixed point \( X^{*} \). At this point, the rates of change
of all the variables in the system are exactly zero, i.e., the l.h.s.'s of the
\( N \) equations in Eq.(5) are zero. The system is said to be in the steady
state at the fixed point. At this point, the state of the system remains unchanged
as a function of time. The only way of changing the state of the system is to
apply perturbations to it. A fixed point is stable if small perturbations around
the point eventually damp out. The stable fixed point acts as an attractor to
the states in its vicinity. The corresponding region in the state space is called
the basin of attraction. The nonlinear dynamics may give rise to more than one
fixed point. If there are two stable fixed points, the system is bistable, i.e.,
two stable steady states are possible. One can similarly define multistability.

The other long-term possibilities for the trajectory in the state space are
a limit cycle and a strange attractor. In the first case, the trajectory goes
towards a closed loop and eventually circulates around it forever. In physical
terms, this corresponds to stable oscillations in the system. The strange attractor
is a set of states to which the trajectory is confined, never stopping or repeating.
Such aperiodic motion is often indicative of chaos in the system. We now discuss
the role of the control parameters R (Eqs. (4) and (5)) in the nonlinear dynamics
of a system. By varying these parameters, one can bring about changes in the
qualitative structure of the dynamics. Such changes are known as bifurcations.
For example, as a parameter is changed, a steady state can become unstable and
replaced by stable oscillations. A system with one stable steady state changes
over to multistability, i.e., the system can exist in multiple steady states.
To give a simple example of bifurcation, consider the rate equation

\begin{equation}
\label{12}
\frac{dx}{dt}=\mu x-x^{2}
\end{equation}
 There are two fixed points of this equation: \( x^{*} \) = 0 and \( x^{*}=\mu  \).
To determine the stability of the fixed points, one undertakes what is known
as the linear stability analysis. One determines the time evolution of a small
perturbation \( \delta x(t)(=x(t)-x^{*}) \) around the fixed point. By substituting
\( x(t)=x^{*}+\delta x(t) \) in Eq.(12) and ignoring terms of the order of
\( (\delta x(t))^{2} \) , one obtains \( \delta x(t)\sim e^{\mu t} \) when
\( x^{*} \) = 0. The fixed point is stable if \( \mu <0 \) since \( \delta x(t) \)
reduces to zero during time evolution. The fixed point is unstable if \( \mu  \)
> 0 and \( \mu _{c} \) = 0 is the bifurcation point. If \( x^{*}=\mu  \),
then \( \delta x(t)\sim e^{-\mu t} \) . Hence the fixed point is unstable if
\( \mu  \) < 0 and stable for \( \mu  \) > 0. Different types of bifurcation
are possible a detailed discussion of which is given in standard textbooks and
reviews \cite{27,28,29} on nonlinear dynamics.

If there is more than one stable fixed point, a switch-like behaviour is possible.
In the case of bistability, the system remains in one stable state until a sufficiently
large perturbation drives the system to the other stable state. The system continues
to remain in the latter state even after the perturbation is removed. The \( \lambda  \)-phage
lysis-lysogeny network offers an example of bistability \cite{9}. The lytic
and the lysogenic states are the two possible steady states. A transition from
the lysogenic to the lytic state occurs on irradiating with ultra-violet light.
In a gene regulatory network, a negative (positive) feedback implies that a
gene product inhibits (promotes) its own level of activity. To give an example,
a protein which represses the transcription of its own gene operates through
negative feedback. It has been found that negative (positive) feedback increases
(decreases) stability in gene regulatory systems \cite{30}. Real life gene
regulatory networks are often complex. Some of the examples are the \( \lambda  \)-phage
lysis-lysogeny circuit, the regulatory network for the activation of the tumour-suppressor
protein p53 \cite{31} and the bacteriophage T7 (another lytic phage which infects
E.coli ) network \cite{32}. Computational modelling studies of these networks
have been undertaken with a view to explain experimental results. The quantitative
agreement between theory and experiment is most often not good. The reasons
are two fold: the complex nature of the networks and the difficulty in carrying
out actual experiments on them. Computational as well as mathematical modelling
of simpler networks is more extensive. Such networks incorporate the essential
features of their more complex counterparts. The models seek to explain experimental
results at a qualitative level. There are also abstract mathematical models
of gene expression/regulation which highlight the general principles and their
outcomes. There are already some good reviews and books on the computational
and mathematical modelling of gene regulatory networks \cite{9,33,34}. For
the purpose of this review, we pick on just one example, that of a synthetic
gene regulatory network which illustrates the importance of nonlinearity in
the dynamics of the network. 

Gardner et al \cite{11} have constructed and tested a synthetic, bistable gene
regulatory network based on the predictions of a simple mathematical model.
The network is called a genetic toggle switch and consists of two repressors
(proteins) and two promoters. The enzyme RNA polymerase (RNAP) binds to the
promoter region of a DNA sequence to initiate the process of transcription.
The initial binding of RNAP to a promoter can be prevented by the binding of
a regulatory protein to an overlapping segment of DNA, called operator. The
gene expression is off in this case. Fig. 2 shows a simple sketch of the toggle
network. The two promoters are designated as \( P_{L} \) and \( Ptrc-2 \)
. \( P_{L} \) drives the expression of the \( lacI \) gene and \( Ptrc-2 \)
that of the \( cI \) gene. The \( lacI \) and \( cI \) genes express the
proteins of the same names. The proteins mutually inhibit the production of
each other, hence the name repressor. The \( lacI \) proteins form tetramers
and the tetramer binds to operator sites adjacent to the \( Ptrc-2 \) promoter,
blocking the transcription of the \( cI \) gene in the process. The \( cI \)
proteins, when produced, form dimers. The repressor dimer cooperatively binds
to the operator sites in the vicinity of the \( P_{L} \) promoter. As a result,
transcription of the \( lacI \) gene is not possible.

The nonlinear dynamics of the toggle network are governed by the following two
equations:

\begin{equation}
\label{13}
\frac{dU}{dt}=\frac{\alpha _{1}}{1+V^{\beta }}-U
\end{equation}
 
\begin{equation}
\label{14}
\frac{dV}{dt}=\frac{\alpha _{2}}{1+U^{\gamma }}-V
\end{equation}
 where \( U \) and \( V \) are the concentrations of \( lacI \) and \( cI \)
proteins respectively, \( \alpha _{1} \) and \( \alpha _{2} \) are the effective
rates of synthesis of \( lacI \) and \( cI \) proteins, \( \beta  \) is the
cooperativity of repression of the \( P_{L} \) promoter and \( \gamma  \)
the same in the case of the \( Ptrc-2 \) promoter. Fig. 3 reveals the origin
of bistability in the system. The nullclines \( \frac{dU}{dt} \) = 0 and \( \frac{dV}{dt} \)
= 0 intersect at three points. These are the fixed points (steady states) of
the dynamics. Two of the fixed points are stable and the third unstable. The
bistability occurs provided \( \beta ,\gamma  \) > 1 (cooperative repression
of transcription) and the rates of synthesis of the two repressors are balanced.
If the rates are not balanced, the nullclines intersect at a single point giving
rise to a single stable steady state (monostability).

In the region of bistability, the two stable steady states correspond to (1)
State 1 (high \( V \) / low \( U \) ) and (2) State 2 ( low \( V \) / high
\( U \) ) respectively. There are two basins of attraction, one above the separatrix
and the other below it. In the \( log(\alpha _{1}) \) vs. \( log(\alpha _{2}) \)
parameter space, bifurcation lines separate the monostable and bistable regions
\cite{11}. The size of the bistable region decreases on reducing the cooperativity
of repression (\( \beta  \) and \( \gamma  \) ). The parameters \( \alpha _{1},\alpha _{2},\beta  \)
and \( \gamma  \) act as the control parameter R changing which a transition
(bifurcation) between monostability and bistability occurs.

In the region of bistability, the toggle is flipped between the stable states
(States 1 and 2) using transient chemical or thermal induction. The chemical
agent isopropyl-\( \beta  \) -D-thiogalactopyranoside (IPTG) can bind to \( lacI \)
tetramers. As a result, the latter cannot bind to the operator region in the
neighbourhood of the promoter \( Ptrc-2 \), i.e., \( lacI \) can no longer
repress the production of the \( cI \) proteins. Suppose the bistable system
is originally in the stable State 2 (high \( U \) ( \( lacI \) ), low \( V \)
( \( cI \) )). On the induction of IPTG, the concentration of the \( cI \)
proteins increases as a function of time. The \( cI \) proteins in their turn
repress the production of \( lacI \) proteins the concentration of which begins
to fall. The dynamics ultimately leads the system to the other fixed point (State
1). The system remains in this stable steady state (low \( U \) / high \( V \)
) even after the removal of the IPTG stimulus. How can the toggle flip back
to State 2 ? This is achieved by using a temperature-sensitive \( cI \) protein
in the network. The degradation rate of this protein increases as temperature
is raised. On raising the temperature to \( 42^{\circ }C \) (actual experiment),
the concentration of \( cI \) proteins starts to fall. Since repression is
less, the concentration of \( lacI \) proteins starts to go up.

The system finally reaches the fixed point corresponding to the stable steady
State 2. After the steady state is reached, the temperature of the system is
reduced (\( 32^{\circ }C \) in the experiment). The system continues to remain
in the steady State 2. A full cycle of the switching process is now completed.
The actual construction of the toggle switch has been accomplished in E.coli
using the standard tools of molecular biology \cite{11}. There is a reasonable
agreement between the theoretical predictions based on Eqs.(13) and (14) and
the results obtained from experiments on the synthetic toggle network. The design
of the network relies significantly on theoretical inputs like identification
of the region of bistability, increasing the cooperativity in repression (\( \beta  \)
and \( \gamma  \) ) to achieve bistability over a wider region in parameter
space etc. As a practical device, the toggle switch may have applications in
biotechnology, biocomputing and gene therapy. As a cellular memory unit, the
toggle provides the basis for ``genetic applets'' which are self-contained,
programmable synthetic gene networks used in the control of cell functions.
In parallel with the toggle work, another synthetic network, the repressilator
has been designed and tested \cite{35}. The repressilator dynamics is again
nonlinear and give rise to oscillations in the concentrations of the cellular
proteins. The design of the network is based on a simple mathematical model
of transcriptional regulation. The repressilator provides insight about the
design principles of other oscillatory systems such as circadian clocks found
in many organisms including cyanobacteria. The genetic toggle switch and the
repressilator demonstrate that theoretical models can provide the design criteria
for the actual construction of synthetic, gene regulatory networks. These simple
networks have applications as practical devices and also help us to understand
the functional properties of the more complex, naturally-occurring networks.

Nonlinear dynamics can give rise to various types of instability one of which
is the Turing instability. In 1952, Turing \cite{36} in a seminal paper proposed
a mechanism for pattern formation in biological systems as well as the development
of structure during the growth of an organism. Examples of biological patterns
are the spots on the skin of a leopard, the stripes of a zebra, the arrangement
of veins on the leaves of a tree etc. Structure formation is initiated by the
process of cell differentiation, an example is the emergence of limbs in an
organism during the growth of the organism from the featureless embryonic stage.
The Turing mechanism involves both reaction as well as diffusion processes.
To illustrate the mechanism, consider two chemical agents, the activator and
the inhibitor. The activator is autocatalytic, i.e., it promotes its own production
as well as that of the inhibitor. The inhibitor, as the name implies, is antagonistic
to the activator and represses its production. Both the chemicals can diffuse
but the inhibitor has a much larger diffusion coefficient. Consider a homogeneous
distribution of the activator and the inhibitor in the system. Increase the
concentration of the activator by a small amount in a local region. This gives
rise to further increases in the local concentrations of the activator and the
inhibitor. The inhibitor quickly diffuses to the surrounding region and prevents
the activator from reaching there. Thus, in the steady state, islands of high
activator concentration exist in a sea of high inhibitor concentration. The
islands constitute what is known as the Turing pattern. Diffusion in general
smooths out concentration differences in a system but the Turing process involving
both reaction and diffusion gives rise to a steady pattern of concentration
gradients. There is now increasing evidence that chemical gradients play a crucial
role in the formation of patterns and cell differentiation in biological systems.
To give an example, the protein bicoid has been found to have a graded concentration
distribution in the Drosophila melanogaster embryo. It is responsible for the
organization of the anterior half of the fly and has been fully characterised
\cite{37,38}. Many reaction-diffusion (RD) models have been proposed based
on the Turing mechanism and some of these can reproduce the patterns observed
in nature \cite{39,40,41,42}. The basic scale of a pattern is larger than the
size of an individual cell and so the RD processes involve more than one cell.
Cells possibly choose developmental pathways depending upon their location in
the concentration gradient. Position-dependent activation of genetic swtches
in the cells may constitute an important step in both pattern and structure
formation. Direct evidence for this in terms of a detailed characterization
of the genes involved and an identification of the actual biochemical reactions
occurring in the cells, is, however, yet to be obtained. Turing patterns have
so far been experimentally observed in certain chemical RD systems in the laboratory
\cite{43,44} and also in some biological systems \cite{45,46}.

\section{Effect of stochasticity}

As already mentioned in the Introduction, stochastic fluctuations in the dynamics
of the gene regulatory network lead to a probabilistic selection of developmental
pathways. The \( \lambda  \)-phage lysis-lysogeny network \cite{47} was discussed
as an example. Fig. 4 shows some of the key components of the network. The complexity
of the full network is captured in Figure 1 of Ref. \cite{13}. We confine our
attention to the simpler network. It consists of two \( \lambda  \) -phage
genes \( cI \) and \( cro \). The corresponding promoters are \( P_{RM} \)
and \( P_{R} \) respectively. Transcription of the gene \( cI \) ( \( cro \)
) expresses the regulatory protein \( \lambda  \) repressor ( \( Cro \) ).
Both the proteins are capable of binding to the operator regions \( O_{R}1 \)
, \( O_{R}2 \) and \( O_{R}3 \). They act antagonistically to control promoter
activity. Transcription of the \( cI \) gene, initiated from the promoter \( P_{RM} \)
, takes place whenever there is no protein of either type binding to \( O_{R}3 \)
. The \( \lambda  \) repressor molecule has a dumbell shape and there is a
tendency for two such molecules to bind and form a dimer. The operator region
\( O_{R}1 \) has the highest affinity for the binding of \( \lambda  \) repressor
dimer. The binding increases the affinity of \( O_{R}2 \) for a second repressor
dimer, i.e., a cooperative binding of dimers to the operator regions \( O_{R}1 \)
and \( O_{R}2 \) takes place. The \( \lambda  \) repressor has both negative
and positive control. If the \( \lambda  \) repressor is present at \( O_{R}2 \)
, transcription of the \( cro \) gene is not possible. This is because the
repressor covers part of the \( DNA \) that a \( RNAP \) molecule must have
access to in order to recognize the promoter \( P_{R} \) , bind to it and initiate
the transcription of the \( cro \) gene. The same repressor at \( O_{R}2 \)
exhibits positive control in helping a RNAP molecule to bind to the promoter
\( P_{RM} \) and begin transcription of the \( cI \) gene. The increase in
the transcription rate is approximately tenfold \cite{47}. If \( O_{R}2 \)
is not occupied by the repressor, the transcription rate of the \( cI \) gene
is low. The reason for the dramatic increase in the transcription rate is the
following. The presence of a repressor dimer bound to \( O_{R}2 \) leads to
an increased affinity of \( P_{RM} \) for \( RNAP \) because the polymerase
is held at \( P_{RM} \) not only by its contacts with the \( DNA \) but also
due to the protein-protein contact with the repressor. In summary, a repressor
dimer bound to \( O_{R}2 \) , represses transcription from \( P_{R} \) but
promotes transcription at \( P_{RM} \) . 

The \( cro \) gene is transcribed only when the operator region \( O_{R}3 \)
is either empty or has \( Cro \) dimer bound to it. The transcription of the
\( cI \) gene cannot take place if the \( O_{R}1 \) and \( O_{R}2 \) regions
are occupied by either protein, \( \lambda  \) repressor and \( Cro \). In
the lysogenic state, all the phage genes are off except for one gene \( cI \)
which produces the protein \( \lambda  \) repressor. The protein in turn binds
to the operators \( O_{R}1 \) and \( O_{R}2 \) in the form of dimers and activates
the transcription of its own gene at \( P_{RM} \). The bound \( \lambda  \)
repressor dimers further prevent transcription initiation at \( P_{R} \) .
Irradiation of the lysogen with ultra-violet light inactivates \( \lambda  \)
repressor making the synthesis of the second regulatory protein \( Cro \) possible.
\( Cro \) promotes lytic growth and competes with the \( \lambda  \) repressor
in occupying the same operator sites. Increased \( Cro \) production leads
to a greater probability of \( Cro \) binding at \( O_{R}3 \) which prevents
the initiation of transcription at \( P_{RM} \) . The concentration of the
\( \lambda  \) repressor starts to fall down as a result. The concentration
of \( Cro \) proteins increases and when it reaches a level such that the operator
regions \( O_{R}1 \) and \( O_{R}2 \) begin to be occupied, the transcription
at \( P_{R} \) is also halted. The switchover from the lysogenic to the lytic
state is further possible by \( recA \) -mediated degradation of the \( \lambda  \)
repressor ( \( recA \) is a catalytic protein ). 

Arkin et al \cite{13} have analysed the stochastic kinetics of the full \( \lambda  \)-phage
network which consists of more genes and regulatory elements than shown in Figure
4. Their detailed investigations show that fluctuations in the rates of gene
expression give rise to random patterns of protein production in individual
cells and wide diversity in instantaneous protein concentrations across cell
populations. Each cell has two developmental pathways: lytic and lysogenic.
The pathway selection depends upon which protein, \( \lambda  \) repressor
or \( Cro \), takes control of the operator region. If it is the \( \lambda  \)
repressor, the lysogenic pathway is chosen. If the \( Cro \) takes control,
the lytic pathway is selected. Due to stochastic fluctuations, the concentrations
of \( \lambda  \) repressor and \( Cro \) vary considerably from cell to cell
tipping the balance in favour of one or the other pathway. As a result, initially
homogeneous cell populations can partition randomly into distinct lytic and
lysogenic subpopulations. Arkin et al have constructed a stochastic kinetic
model of the \( \lambda  \)-phage circuit and based on model calculations predicted
the fraction of infected cells selecting the lysogenic pathway at different
phage:cell ratios. The theoretical results are consistent with the experimental
results of Kourilsky \cite{48}. The kinetic model uses the stochastic formulation
of chemical kinetics \cite{14,15}, stochastic mechanisms of gene expression
\cite{12} and a statistical-thermodynamical model of promoter regulation \cite{49}.
Probabilistic selection of developmental pathways occurs in several other gene
regulatory networks producing stochastic phenotypic outcomes. Some examples
are given in Table 4 of Ref. \cite{10}.

We now describe the well-known Gillespie algorithm \cite{14,15} which is incresingly
being used by biologists in the stochastic kinetic approach to the study of
gene expression and regulation in different systems Let us consider a system
of \( N \) chemicals participating in \( M \) reactions \( R_{\mu } \). The
state of the system at any instant of time \( t \) is represented as \( (X_{1},...,X_{N}) \)
where \( X_{i} \) is the number of molecules of the ith chemical species. Two
questions have to be answered to determine how the system evolves in time: (1)
when will the next reaction occur and (2) what type of reaction will it be ?
Let 

\( C_{\mu }dt \) = the probability that an \( R_{\mu } \) \( (\mu =1,...,M) \)
reaction occurs in the next infinitesimal time interval \( dt \) for a particular
combination of the reactant molecules. Let \( h_{\mu } \) be the number of
distinct combinations of molecules available in the state \( (X_{1},...X_{N}) \)
for the \( R_{\mu } \) reaction.\\
As an example, consider the reaction

\begin{equation}
\label{15}
A+B\rightarrow C
\end{equation}
 Let \( X_{1} \) and \( X_{2} \) be the number of molecules of types A and
B respectively. Then \( h=X_{1}X_{2} \) . Let

\( a_{\mu }dt= \) \( h_{\mu }C_{\mu }dt \) be the probability that an \( R_{\mu } \)
reaction occurs in time \( (t,t+dt) \) given the system is in the state \( (X_{1},...,X_{N}) \)
at time \( t \). \\
The reaction probability density function \( P(\tau ,\mu )d\tau  \) is the
probability that given the state \( (X_{1},...,X_{N}) \) at time \( t \),
the next reaction will occur in the infinitesimal time interval \( (t+\tau ,t+\tau +d\tau ) \)
and will be an \( R_{\mu } \) reaction,

\begin{equation}
\label{16}
P(\tau ,\mu )d\tau =P_{0}(\tau )a_{\mu }d\tau 
\end{equation}
 where \( P_{0}(\tau ) \) is the probability that no reaction occurs in the
time interval \( (t,t+\tau ) \) and \( a_{\mu }d\tau  \) is the subsequent
probability that an \( R_{\mu } \) reaction occurs in the time interval \( (t+\tau ,t+\tau +d\tau ) \).
Now

\begin{equation}
\label{17}
P_{0}(\tau +d\tau )=P_{0}(\tau )\left[ 1-\sum ^{M}_{\nu =1}a_{\nu }d\tau \right] 
\end{equation}
 where the expression inside the bracket is the probability that no reaction
occurs in time \( d\tau  \) from the state \( (X_{1},....,X_{N}) \). Eq. (17)
can be solved to obtain

\begin{equation}
\label{18}
P_{0}(\tau )=exp\left[ -\sum ^{M}_{\nu =1}a_{\nu }\tau \right] 
\end{equation}
 Substituting for \( P_{0}(\tau ) \) in Eq. (16), one gets

\begin{equation}
\label{19 }
P(\tau ,\mu )=a_{\mu }exp\left( -a_{0}\tau \right) 
\end{equation}
 if \( 0\leq \tau <\infty  \), \( \mu =1,...,M \) and \( P(\tau ,\mu )=0 \)
otherwise.

\begin{equation}
\label{20}
a_{\mu }=h_{\mu }C_{\mu },(\mu =1,...,M)
\end{equation}
 and

\begin{equation}
\label{21}
a_{0}=\sum ^{M}_{\nu =1}a_{\nu }
\end{equation}
 Now the goal is to generate a pair of random numbers \( (\tau ,\mu ) \) acording
to the probability distribution (19). To do this, use the standard random number
generator to obtain two random numbers from the uniform distribution in the
unit interval. Take 

\begin{equation}
\label{22}
\tau =\frac{1}{a_{0}}ln\left( \frac{1}{r_{1}}\right) 
\end{equation}
 and \( \mu  \) is chosen to be the integer for which

\begin{equation}
\label{23}
\sum ^{\mu -1}_{\nu =1}a_{\nu }<r_{2}a_{0}\leq \sum ^{\mu }_{\nu =1}a_{\nu }
\end{equation}
 The pair of numbers \( (\tau ,\mu ) \), (Eqs. (22) and (23)), belongs to the
set of random pairs described by the probability density function \( P(\tau ,\mu ) \).
For a rigorous proof of this see Refs. \cite{14,15}. Once \( (\tau ,\mu ) \)
are known, put 

\begin{equation}
\label{24}
t=t+\tau 
\end{equation}
 and adjust the \( X_{i} \) values according to the \( R_{\mu } \) reaction.
If the \( R_{\mu } \) reaction is the one shown in Eq.(15), both \( X_{1} \)
and \( X_{2} \) have to be decreased by 1 and \( X_{3}, \) the number of molecules
of C, increased by 1.

The input values at time \( t \) = 0 are \( h_{\nu },C_{\nu }(\nu =1,...M) \)
and the initial values of \( X_{i}(i=1,...,N) \). The steps of the Gillespie
algorithm are:

Step 1\\
Calculate \( a_{\nu }=h_{\nu }C_{\nu }(\nu =1,...,M) \) and \( a_{0}=\sum ^{M}_{\nu =1}a_{\nu } \).

Step 2\\
Generate \( r_{1} \) and \( r_{2} \) with the help of a uniform random number
generator. Calculate \( \tau  \) and \( \mu  \) according to the formulae
in Eqs. (22) and (23). 

Step 3\\
Advance \( t \) by \( \tau  \) (Eq.(24)) and adjust the \( X_{i} \) values
according to \( R_{\mu } \). Then repeat the steps from Step 1 to further advance
the system in time.

Recently, Kierzek et \cite{50} al have used the Gillespie algorithm to study
a stochastic kinetic model of prokaryotic gene expression. They explicitly considered
ten biochemical reactions:

\begin{equation}
\label{25}
P+RNAP\rightarrow P_{-}RNAP
\end{equation}
 
\begin{equation}
\label{26}
P_{-}RNAP\rightarrow P+RNAP
\end{equation}

\begin{equation}
\label{27}
P_{-}RNAP\rightarrow TrRNAP
\end{equation}

\begin{equation}
\label{28}
TrRNAP\rightarrow RBS+P+EIRNAP
\end{equation}
 where \( P \) denotes the promoter region of the gene and \( P_{-}RNAP \)
the bound promoter-\( RNAP \) complex. Reaction 2 (Eq.(26)) describes \( RNAP \)
dissociation and Reaction 3 the isomerization of ``closed complex'' to ``open
complex'', \( TrRNAP \) is the activated  \( RNAP- \)promoter complex. Reaction
4 describes clearance of promoter region by \( RNAP, \) \( EIRNAP \) stands
for \( RNAP \) transcribing the gene and synthesizing the \( mRNA \) molecule
and \( RBS \) is the ribosome binding site on \( mRNA \). The other reactions
are:

\begin{equation}
\label{29}
Ribosome+RBS\rightarrow RibRBS
\end{equation}
 
\begin{equation}
\label{30}
RibRBS\rightarrow RBS+Ribosome
\end{equation}
 
\begin{equation}
\label{31}
RibRBS\rightarrow EIRib+RBS
\end{equation}
 
\begin{equation}
\label{32}
RBS\rightarrow decay
\end{equation}
 
\begin{equation}
\label{33}
EIRib\rightarrow protein
\end{equation}
 
\begin{equation}
\label{34}
Protein\rightarrow decay
\end{equation}
 Reactions 5-10 (Eqs. (29)-(34)) describe translation, \( mRNA \) decay and
protein degradation. Reaction 5 describes \( Ribosome \) binding to \( RBS \)
, the bound complex is designated as \( RibRBS \) . Reaction 6 is the dissociation
of the bound complex. Reaction 7 describes \( Ribosome \) binding site clearance,
\( EIRib \) is the \( Ribosome \) which translates the \( mRNA \). Reaction
8 has degradation of \( RBS \) by the enzyme \( RNAaseE \). \( RNAaseE \)
and \( Ribosomes \) are in competition to occupy \( RBS \). If \( RNAaseE \)
binds first then it initiates the degradation of \( mRNA \) but does not interfere
with the movement of the already bound \( Ribosomes \) engaged in the process
of translation. Every \( Ribosome \) which successfully binds to the \( RBS \)
completes translation of the protein. Reaction 9 corresponds to the completion
of protein synthesis. Reaction 10 represents protein decay. The stochastic rate
constants \( C_{\mu }'s \) of the different reactions, needed as inputs to
the Gillespie algorithm, can be calculated from the more familiar chemical rate
constants listed in Kierzek et al's paper \cite{50}. For first order chemical
reactions, the stochastic rate constant is equal to the rate constant of a chemical
reaction. For second order reactions, the stochastic rate constant is equal
to the rate constant divided by the volume of the system (in this case a cell). 

In the last part of this Section, we describe a cooperative stochastic model
of gene expression proposed by us \cite{20}. As already explained in the Introduction,
the model has been constructed to explain the bimodal distribution in gene expression
observed in recent experiments. The model describes the transcription of a single
gene with one promoter region. There is one operaor region to which a regulatory
protein \( R \) can bind. This prevents the binding of a \( RNAP \) to the
promoter so that transcription of the gene cannot be initiated. There is a finite
probability that the bound \( R \) molecule dissociates from the operator at
any instant of time. \( RNAP \) molecule then has a certain probability of
binding to the promoter and initiating transcription.

Each of the possibilities described above actually involves a series of physico-chemical
processes, a detailed characterization of which is not required for the model
of gene expression proposed by us. We represent a gene by a one-dimensional
lattice of \( n+2 \) sites. The first two sites represent the operator and
promoter respectively. The lattice is a coarse-grained description of an actual
gene. In reality, the operator and promoter regions may extend over a certain
number of base pairs in the DNA and they can be overlapping or not. In our model,
they are represented as single sites. Each of the other sites in the lattice
represents a finite number of base pairs in the DNA molecule.

The different physico-chemical processes are lumped together into a few simple
events which are random in nature. This lumping together avoids unnecessary
complexity that has no bearing on the basic nature of the process. The operator
\( (O) \) and the promoter \( (P) \) together can be in four possible configurations:
\( 10,01,00 \) and \( 11 \). The numbers ``\( 1 \)'' and ``\( 0 \)''
stand for ``occupied'' and ``unoccupied''. The configuration \( ij \) describes
the occupation status of \( O \) \( (i) \) and \( P \) \( (j) \). For example,
the configuration \( 10 \) corresponds to \( O \) being occupied by a \( R \)
molecule and \( P \) being unoccupied. Similarly, in the configuration \( 01 \),
\( O \) is unoccupied and \( P \) is occupied by a \( RNAP \) molecule. Binding
of \( R \) and \( RNAP \) molecules are mutually exclusive so that the configuration
\( 11 \) is strictly prohibited. Given a \( 00 \) configuration at time \( t \),
the transition probabilities to configurations \( 10 \) and \( 01 \) at time
\( t+1 \) are \( p_{1} \) and \( p_{2} \) respectively. The probability of
remaining in the configuration \( 00 \) is \( 1-p_{1}-p_{2} \). A \( 10 \)
configuration at time \( t \) goes to a \( 00 \) configuration at time \( t+1 \)
with probability \( p_{3} \) and remains unchanged with probability \( 1-p_{3} \).
We have assumed all the probabilities to be time-independent. The \( RNAP \)
molecule once bound to the promoter initiates transcription in the next time
step, i.e., the \( 01 \) configuration makes a transition to a \( 00 \) configuration
with probability \( 1 \). The motion of \( RNAP \) is in the forward direction
and the molecule covers a unit distance (the distance between two successive
lattice sites) in each time step. Once the molecule reaches the last site of
the lattice, the transcription ends and a \( mRNA \) is synthesized.

The second major feature of our model is the cooperative binding of \( RNAP \)
to the promoter, when an adjacent \( RNAP \) molecule is present. This implies
that there is a higher probability of binding of \( RNAP \) to the promoter
in one time step if another \( RNAP \) molecule is present at the site next
to the promoter. In our model, the probability of cooperative binding is \( p_{4} \)
which is larger than \( p_{2} \). The probabilities \( p_{1} \) and \( 1-p_{1}-p_{2} \)
are changed to new values \( p_{5} \) and \( 1-p_{4}-p_{5} \) respectively.
Degradation of \( mRNA \) is taken into account by assuming the decay rate
to be given by \( \mu N \), where \( N \) is the number of \( mRNAs \) present
at time \( t \). The number of \( mRNAs \) produced as a function of time
is studied by Monte Carlo simulation. For the sake of simplicity, we have not
tried to simulate protein levels or enzymatic products thereof, i.e., we study
gene expression upto the level of transcription (\( mRNA \) synthesis). Since
the number of protein molecules and converted products should be proportional
to the number of \( mRNA \) molecules, no loss of generality is introduced
by this simplification. The lattice consists of 52 sites \( (n=50) \). Stochastic
events are simulated with the help of a random number generator. The updating
rule of our cellular automaton (CA) model is that in each time step \( t \),
the occupation status ( \( 0 \) or \( 1 \)) of each site (except for the \( O \)
site) at time \( t-1 \) is transferred to the nearest-neighbour site towards
the right. If the last site is \( 1 \) at \( t-1 \), a \( mRNA \) is synthesized
at \( t \) and the number of \( mRNAs \) increases by one. In the same time
step, the configuration \( ij \) of \( OP \) is determined with the probabilities
already specified. Thus, in each time step, the \( RNAP \) molecule, if present
on the gene, moves forward by unit lattice distance (progression of transcription)
followed by the updating of the \( OP \) configuration. Figure 5 shows the
concentration \( [mRNA] \) of \( mRNA \) molecules in the cell as a function
of time for the parameter values \( p_{1}=0.5,p_{2}=0.5,p_{3}=0.3,p_{4}=0.85,p_{5}=0.05 \)
and \( \mu =0.4 \). Note that an almost four-fold increase in the probability
of \( RNAP \) binding is assumed due to cooperativity. The stochastic nature
of the gene expression is evident from the figure, with random intervals between
the bursts of activity. One also notices the presence of several bursts of large
size. It is important to emphasize that the frequency of transition between
high and low expression levels is a function of the parameter values chosen
and may be low for certain parameter values. For the probability values considered,
the two predominantly favourable states are when the gene expression is off
(state 1) and when a large amount of gene expression takes place (state 2).
In the absence of \( RNAP \) binding, state 1 has greater weight but with the
chance binding of \( RNAP \) to the promoter (probability \( p_{2} \) for
this is small), the weight shifts to state 2 until another stochastic event
terminates cooperative binding and the gene reverts to state 1. The probability
of obtaining a train of \( N \) successive transcribing \( RNAP \) molecules
is \( p_{2}p^{N-1}_{4}(1-p_{4}) \). This is the geometric distribution function
and the mean and the variance of the distribution are given by \( \frac{p_{2}}{1-p_{4}} \)
and \( \frac{p_{2}(1+p_{4}-p_{2})}{(1-p_{4})^{2}} \) respectively. For the
probability values already specified, the simulation has been repeated for an
ensemble of 3000 cells. For each cell, the time evolution is upto 10000 time
steps. Figure 6 shows the distribution of the number \( N(m) \) of cells versus
the fraction \( m \) of the maximal number of \( mRNA \) molecules produced
after \( 10000 \) time steps. Two distinct peaks are seen corresponding to
zero and maximal gene expression. Such a bimodal distribution occurs over a
range of parameter values. 

Some theories have been proposed so far to explain the so-called ``all or none''
phenomenon in gene expression \cite{19,51,52}. These theories are mostly based
on an autocatalytic feedback mechanism, synthesis of the gene product gives
rise to the transport or production of an activator molecule. While such processes
are certainly possible, the bimodal distribution is a much more general phenomenon
and has now been found in many types of cells, from bacterial to eukaryotic
and for different types of promoters \cite{16,17,18}. The two major features
of the model of gene expression that we have proposed are stochasticity and
cooperative binding of \( RNAP \). There is by now enough experimental evidence
of stochasticity in gene expression. Our suggestion of cooperative binding of
\( RNAP \) is novel and there is no direct experimental verification of the
proposal as yet. There are some recent experiments which provide indirect evidence
and these are discussed in Ref. \cite{20}.

\section{Concluding remarks}

In this review we have given an elementary introduction to some of the major
aspects of biological networks, namely, topological characteristics, nonlinear
dynamics and the role of stochasticity in gene expression and regulation. The
main aim of the review is to highlight the usefulness of interdisciplinary approaches
in the study of both natural and synthetic biological networks. Some important
features of such networks have not been discussed in the review. One of these
is the operational reliability of networks in spite of randomness in basic regulatory
mechanisms. Many regulatory pathways do have highly predictable outcomes even
when stochastic fluctuations are considerable. Cells adopt various strategies
like populational transcriptional cooperation, checkpoints to ensure that cascaded
events are appropriately synchronised, redundancy and feedback to achieve regulatory
determinism. Some of these ideas are discussed in Ref. \cite{10}. The complexity
of biological networks raises the question of their functional stability. In
particular, the issue of interest is the sensitivity of networks to variations
in their biochemical parameters. Barkai and Leibler \cite{53} have studied
a biochemical network responsible for bacterial chemotaxis and shown that the
functional properties of the network are robust, i.e., relatively insensitive
to changes in biochemical parameters like reaction rate constants and enzymatic
concentrations. Bialek \cite{54} has shown that extremely stable biochemical
switches can be constructed from small numbers of molecules though intuitively
one expects such systems to be prone to instability due to the inherent noise.

Metzler \cite{55} in a recent paper has shown that spatial fluctuations in
the distribution of regulatory molecules play a non-trivial role in genetic
switching processes. Apart from internal stochastic fluctuations, external noise
originating in random variations in the environment or in the externally set
control parameters, may affect the functioning of a biological network. Hasty
et al \cite{56} have proposed a synthetic genetic network in which external
noise is utilised to operate a protein switch (short noise pulses are used to
turn protein production ``on'' and ``off''). In another novel application,
external noise is used to amplify gene expression, i.e., protein production
by a considerable amount.

Genetic networks with many components are difficult to analyze using conventional
techniques. Many parallels have been drawn in the functioning of genetic and
electrical circuits \cite{57,58}. In electrical engineering, there are well
developed techniques of circuit analysis which can be used to characterise the
operation of complex electrical networks. Some of these techniques are increasingly
being used to study genetic networks. Engineers are familiar with some of the
design principles of biological networks. Rapid transitions between the two
stable states of a system can be brought about by positive feedback loops. Negative
feedback loops control the value of an output parameter to be within a narrow
range even if there are wide fluctuations in the input. Coincidence detection
systems activate an output provided two or more events occur simultaneously.
Parallel connections enable a device to remain functional in the event of failures
in one of the lines. One can give analogous examples from biology. One set of
positive feedback loops is responsible for the rapid transition of cells into
mitosis (division of cell nucleus), another set brings about the exit from mitosis
in an irreversible manner. Gene transcription in eukaryotes involve coincidence
detection. A \( mRNA \) can be produced only if the promoters regulating gene
expression are occupied by the different transcription factors. These examples
indicate that general principles govern the functioning of genetic and electrical
networks though there are other aspects of such networks which are not common
to both. Biological networks constitute a field of research the interdisciplinary
nature of which will become more evident as we progress into the twenty first
century.\\
\textbf{Acknowledgement: The Author thanks Subhasis Banerjee for help in drawing
the figures.}

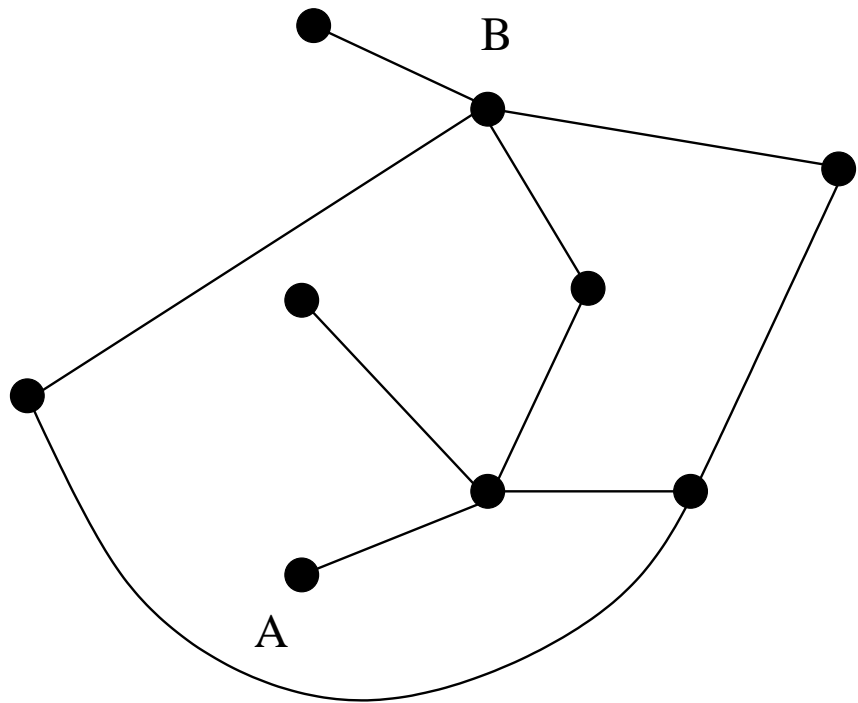
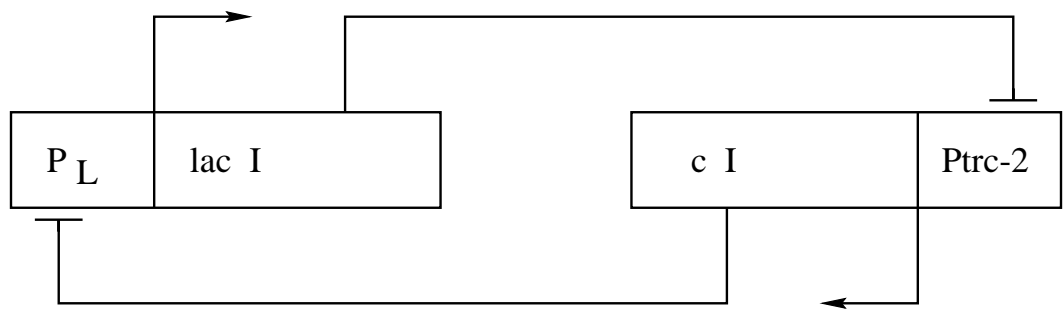
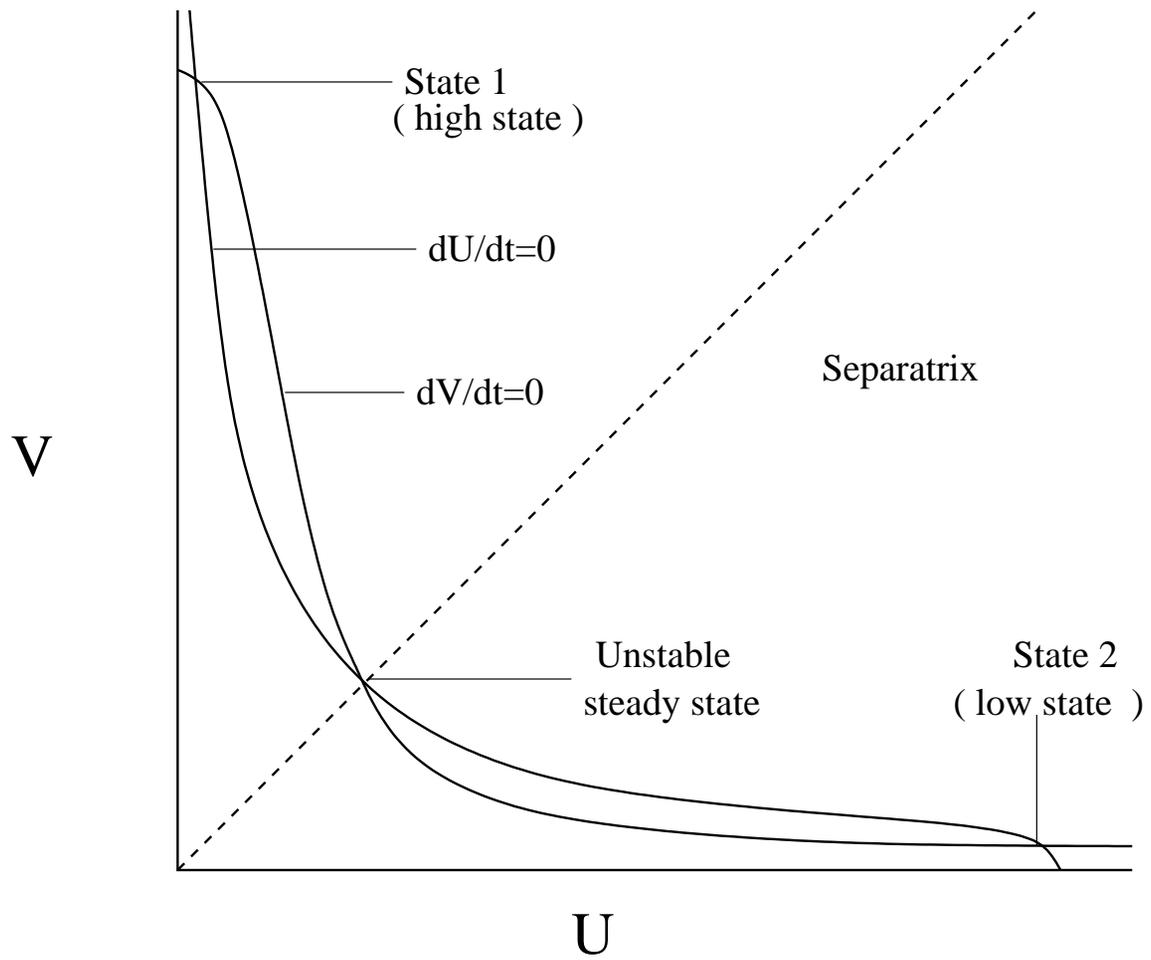
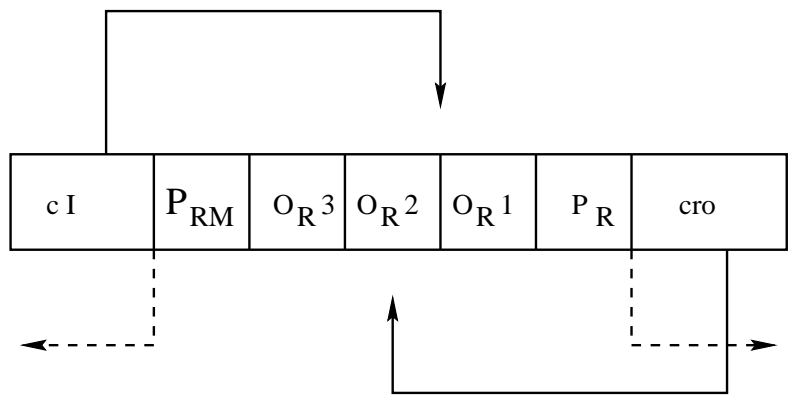
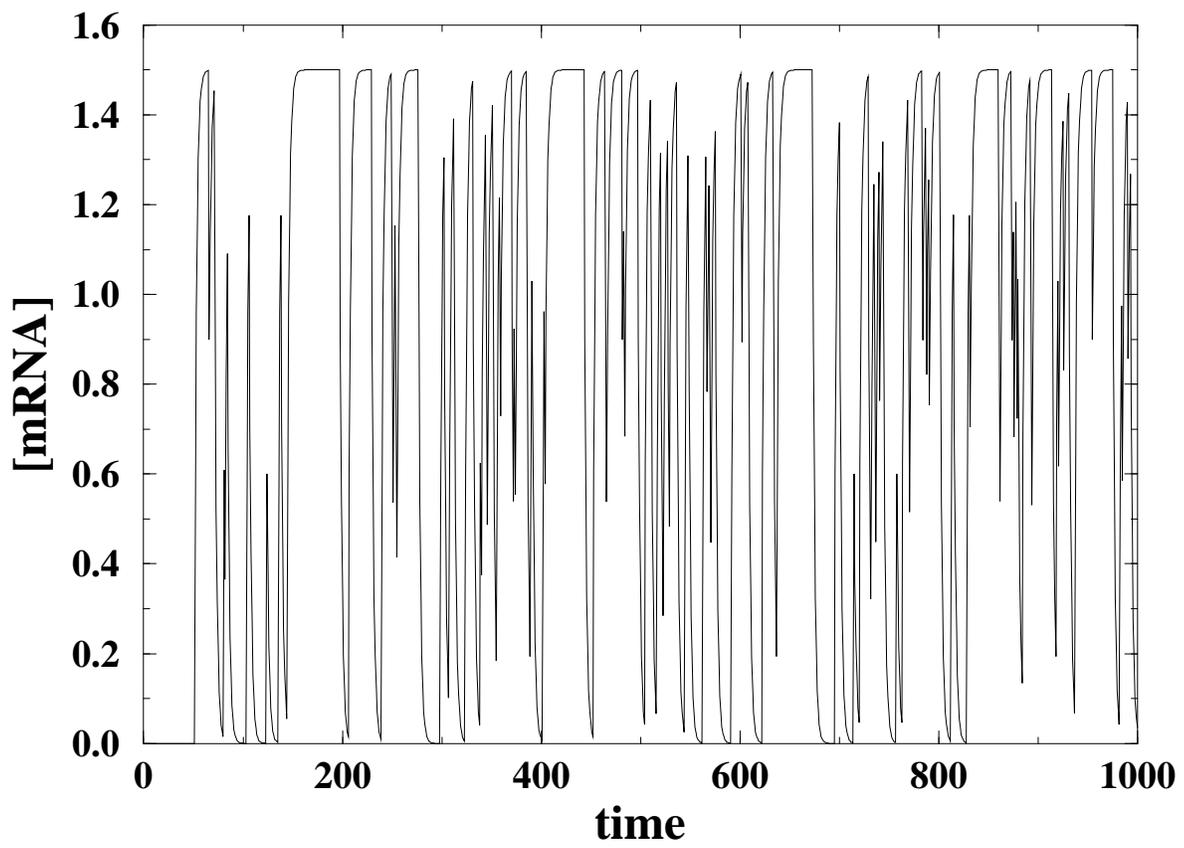
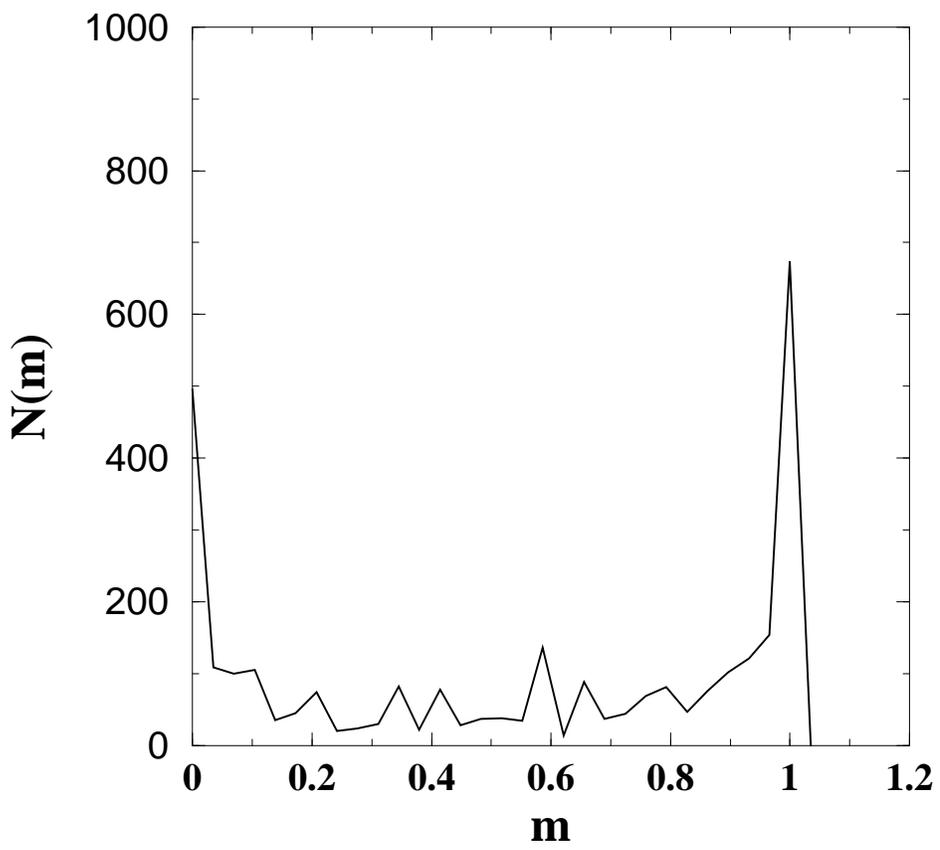
\end{document}